\documentclass[useAMS,usenatbib]{mn2e}
\usepackage{psfig}

\usepackage{amstext}
\usepackage{amssymb}
\usepackage{amsmath, float}
\usepackage{graphicx}
\usepackage{txfonts}
\usepackage{amsmath,mathtools,revsymb,amssymb}
\usepackage{booktabs,multirow,calc,xspace,rotating}
\usepackage{url}

\newcommand       \be           {\begin{equation}}
\newcommand       \ee           {\end{equation}}
\newcommand       \bea          {\begin{eqnarray}}
\newcommand       \eea          {\end{eqnarray}}
\newcommand       \kms		{\,{\rm km \,\, s}^{-1}}

\renewcommand*{\emph}[1]{\textit{\textbf{#1}}}

\begin{document}

\title[Accretion Discs in Failed Supernovae]{Black Hole Accretion Discs and Luminous Transients in Failed Supernovae from Non-Rotating Supergiants}
\author[E. Quataert, D. Lecoanet, \& E. Coughlin]{E. Quataert$^{1}$, D. Lecoanet$^{2}$, \& E. R. Coughlin$^{3,4}$  \\
\\
  $^{1}$Astronomy Department and Theoretical Astrophysics Center, University of California, Berkeley, 601 Campbell Hall, Berkeley CA, 94720 \\
  $^{2}$Princeton Center for Theoretical Science, Princeton University, Princeton, NJ 08544, USA \\
  $^{3}$Department of Physics and Columbia Astrophysics Laboratory, Columbia University, New York, NY 10027, USA \\
  $^{4}$Einstein Fellow}

\maketitle

\begin{abstract}

We show that for supergiants, net angular momentum is not a necessary condition for forming accretion discs during core-collapse.   Even absent net rotation, convective motions in the outer parts of supergiants generate mean horizontal flows at a given radius with velocities of $\sim 1 \kms$; the direction of the mean flow will vary as a function of height through the convection zone.   We confirm these analytic estimates using Cartesian Boussinesq convection simulations.   These mean horizontal flows lead to a random angular momentum in supergiant convection zones that exceeds that of the last stable circular orbit of a black hole by a factor of $\sim 10$.   As a result, failed explosions of supergiants - in which the accretion shock onto the neutron star does not revive, leading to black hole formation - may often produce accretion discs that can power day-week (blue supergiants) or week-year (yellow and red supergiants) non-thermal and thermal transients through winds and jets.   These transients will be especially time variable because the angular momentum of the accreting material will vary substantially in time.       Observed sources such as Swift J1644+57, iPTF14hls, and SN 2018cow, as well as energetic Type II supernovae (OGLE-2014-SN-073) may be produced by this mechanism.

\end{abstract}

\voffset=-2cm

\section{Introduction}
\label{sec:intro}
\vspace{-0.1cm}

The structure of a  massive star prior to core-collapse strongly influences the likelihood of a neutrino-powered supernova (SN) explosion \citep{OConnor_Ott_2011}.   Stars with a more compact core and a shallower density profile outside the core are more difficult to explode because the accretion rate onto the proto-neutron star (PNS) is larger, stifling the explosion \citep{Ertl2016}.  It is thus  possible that a subset of massive stars do not lead to a neutrino-powered SN explosion; this subset may depend non-trivially on mass, metallicity, and/or rotation \citep{Sukhbold2018}.    There are also a variety of  observational hints that some fraction of massive star core-collapse may indeed fail to produce a canonical luminous energetic explosion (see, e.g., \citealt{Kochanek_2015} and references therein).    

Absent angular momentum, the most well understood way to eject mass during a failed SN is that the radiation of $\sim10^{53}$ ergs of neutrino energy during the proto-NS phase creates a nearly-instantaneous decrease in the mass of the core as viewed by the outer layers of the star.
This generates an outgoing sound pulse that steepens into a shock and can unbind some of the stellar envelope (\citealt{Nadezhin_1980,Lovegrove_Woosley_2013}; \citealt{fernandez18}, hereafter F18).   This shock energy is, however, $\lesssim 10^{48}$ ergs \citep{Coughlin18} and can only produce a faint thermally powered transient.   The angular momentum of the progenitor star is thus usually assumed to be the key property that  determines whether  black hole formation in a nominally failed SN might nonetheless produce a luminous transient:  significant angular momentum allows the collapse to tap into black hole accretion energy, thereby turning a dud into a success.       Indeed, accretion discs  produced in failed SNe were one of the original collapsar variants proposed to account for long-duration gamma-ray bursts \citep{MacFadyen_Woosley_1999}.    Related ideas have since seen traction in explaining a  range of other observations, including much longer-duration gamma-ray transients (\citealt{Quataert_Kasen_2012, Woosley_Heger_2012}), some unusual Type II SNe (e.g., \citealt{Moriya2018}) and super-luminous SNe (e.g., \citealt{Dexter_Kasen_2013}).      

In this {\em Letter}, we show that in the collapse of massive supergiants, net angular momentum in the progenitor is not a necessary condition for forming an accretion disc and tapping into accretion energy to power luminous transients.    The convection zones of supergiants have  large random mean horizontal velocities (even without a net angular momentum), so that spherical accretion onto a newly formed black hole cannot occur.   We present the basic physics in \S \ref{sec:flows}, present its application to failed SNe in \S \ref{sec:failedSNe}, and discuss our results and their implications in \S \ref{sec:discussion}.

\vspace{-0.5cm}
\section{Mean Horizontal Convective Flows}
\label{sec:flows}

\subsection{Analytic Estimates}
\label{sec:analytics}

We begin with an order of magnitude estimate of the random mean flows in convection zones, due to the finite number of convective eddies. 
Consider a region of a star at radius $r$ with local scale-height  $H$ and convective velocity  $v_c$; we follow mixing length theory and assume that the typical size of a convective eddy is $\sim H$.   The latter implies that the number of eddies at radius $r$ is $N_{\rm edd} \simeq 4 \pi (r/H)^2$.   
Even if the total horizontal momentum integrated over the convection zone vanishes, the finite number of eddies implies that {at a given radius} there is mean horizontal velocity of
\be
v_h \sim \frac{v_c}{\sqrt{4 \pi}} \frac{H}{r}
\label{eq:vh}
\ee
In a star this corresponds to a random specific angular momentum 
\be
j_{\rm rand} \sim \frac{ H v_c}{\sqrt{4 \pi}} \sim 6 \times 10^{17} \frac{r \, v_c }{10^3 R_\odot \, \kms} \left(\frac{H/r}{0.3}\right) \, {\rm cm^2 \, s^{-1}}
\label{eq:jrms}
\ee
where we have normalized values to be appropriate for the convection zones of supergiants ($H/r \sim 0.3$, $v_c \sim 10 \kms$ and $r \sim 100 R_\odot$, which correspond to mean horizontal flows with $v_h \sim 1 \kms$).   We show below that in Boussinesq convection simulations this estimate is accurate to about a factor of 2.   For comparison, the specific angular momentum of the Inner-Most Stable Circular Orbit (ISCO) for a black hole is given by 
\be
j_{\rm ISCO} = 1.15-3.5 \, \frac{GM}{c} \sim 0.5-1.5 \times 10^{17} \, \left(\frac{M}{10 \, M_\odot}\right) \, {\rm cm^2 \, s^{-1}}
\label{eq:jISCO}
\ee
where the range corresponds to spins $a/M$ ranging from $0-1$.   A comparison of equations \ref{eq:jrms} and \ref{eq:jISCO} shows that the angular momentum associated with random mean flows in the outer convection zones of supergiants can be appreciable, and is likely sufficient to preclude the matter from falling spherically onto the central black hole even absent net angular momentum in the convection zone.   This outer convective envelope is, however, weakly bound, so  it is prone to being ejected by even  low energy explosions (see \S \ref{sec:discussion}).

\vspace{-0.5cm}
\subsection{Simulations of Convection}
\label{sec:sims}

To assess the estimates in equations 1 \& 2 we carry out a number of Cartesian box simulations of convection and study the mean horizontal flows in such simulations. Random horizontal flows in a Cartesian box absent a net horizontal momentum are directly analogous to random angular momentum in a spherical star absent net rotation. The Cartesian box simulations are, however, computationally easier and we believe suffice for this initial study.

Our simulations utilize the Boussinesq approximation, in which density perturbations are neglected except in the buoyancy term \citep[see, e.g.,][and references within]{spiegel60}. The Boussinesq approximation is formally valid when the flow is subsonic and occurs on length-scales smaller than the scale-height. For the stars of interest, the radial extent of the convection zone is multiple scale-heights and the convection becomes nearly sonic in the low density outer layers. Thus, the Boussinesq approximation does not formally hold in all of the regions of interest. Nevertheless,  the Boussinesq approximation is the simplest computational model to test  the mechanism described in \S ~\ref{sec:analytics}.   

Our calculations are for the classic Rayleigh--Benard convection problem, with stress-free boundaries.   This can be non-dimensionalized in terms of a Rayleigh number $Ra=\alpha gH^3/(\nu\kappa)$ and a Prandtl number $Pr=\nu/\kappa$ (which we take to be unity), where $g$ is the (constant) strength of gravity, $\alpha$ is the thermal expansion coefficient, and the momentum and thermal diffusivities are $\nu$ and $\kappa$ respectively. The boundary conditions are fixed temperature 
at the bottom/top, impenetrable, and stress-free to ensure the total horizontal momentum of the layer is zero. The system is assumed to have a height of one scale-height $H$, and a horizontal size of $L_x=L_y=4H$ or $8H$. 


\begin{table}
\caption{Properties of the four simulations considered in this letter. The input parameters are the Rayleigh number $Ra$ and the aspect ratio $L_x/H$ (which equals $L_y/H$); the vertical height is $L_z \equiv H$. We then measure the root mean square velocity of the convection, $v_c$, and the average mean horizontal velocity $v_h$. The simulation is run to $t_{\rm end}$. \label{tab:sims}}
\begin{tabular}{c|cc|cccc}
name  & $Ra$ & $L_x/H$ & $v_c/\sqrt{\alpha gH}$ & $v_h/v_c$ & $t_0H/v_c$ & $t_{\rm end}H/v_c$  \\[3pt] \hline
R6L4 & $10^6$ & 4 & 0.34 & 0.102 & 21 & 68 \\
R6L8 & $10^6$ & 8 & 0.33 & 0.053 & 20 & 132 \\
R8L4 & $10^8$ & 4 & 0.27 & 0.104 & 16 & 66 \\
R8L8 & $10^8$ & 8 & 0.27 & 0.052 & 16 & 33\\
\end{tabular}
\end{table}

We solve 
the  Boussinesq equations with the Dedalus pseudo-spectral code\footnote{More information at \url{dedalus-project.org}.} \citep{burns16}. We represent each variable 
using Fourier series in the horizontal $x$ and $y$ directions, and a series of Chebyshev polynomials in the $z$ direction. Boundary conditions are imposed using the $\tau$ method. We run low resolution simulations with $Ra=10^6$, as well as higher resolution, turbulent simulations with $Ra=10^8$. Table~\ref{tab:sims} has the main properties of each simulation. We discretize the problem by truncating the Fourier series with 32 modes per scale height for simulations with $Ra=10^6$, and 128 modes per scale height for simulations with $Ra=10^8$. In the vertical direction, we truncate the series of Chebyshev polynomials with 64 modes for simulations with $Ra=10^6$ and 256 modes for simulations with $Ra=10^8$. We evaluate nonlinear terms using the pseudo-spectral technique, and use $3/2$ padding dealiasing to prevent aliasing errors. Thus, our most turbulent simulation with $Ra=10^8$ and $L_x=L_y=8H$ has $1536\times1536\times384$ grid points. For timestepping, we use an implicit-explicit second-order semi-backwards differencing formula.

Because our method does not explicitly conserve horizontal momentum, we can use the volume averaged momentum as an estimate of the errors in our simulation. For our highest resolution simulation, the domain-averaged momentum at the end of the simulation is about $10^{-9}$ times the convective velocity.

\begin{figure}
  \centerline{\includegraphics{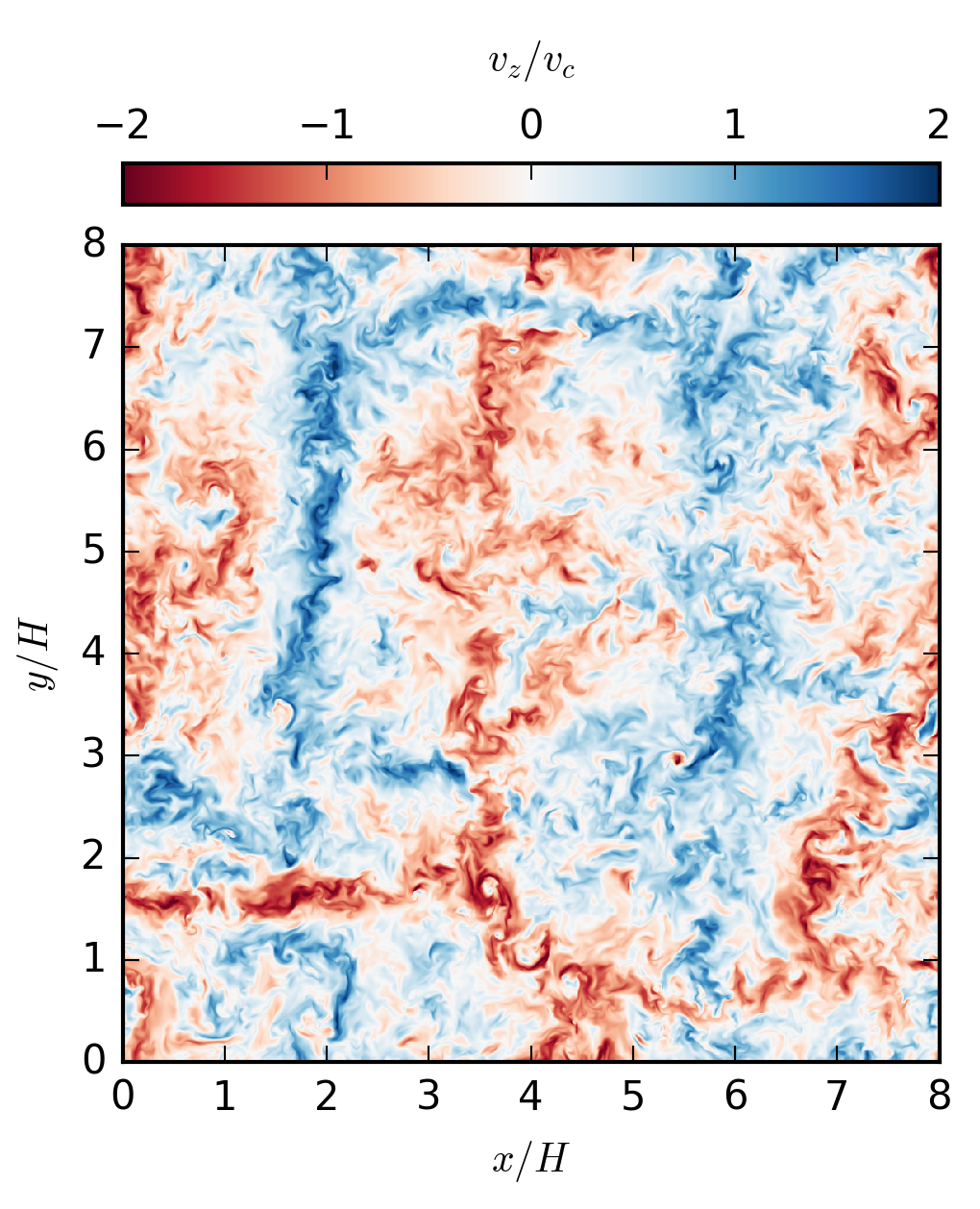}}
  \caption{Vertical velocity at the midplane, at the final time of simulation R8L8. The convection organizes into long, coherent up- and down-flow lanes, each of which has a width of about $H$. Figure~\ref{fig:u_mean} shows the mean horizontal velocity in as a function of time in this simulation.}
\label{fig:w}
\end{figure}

The simulations are 
run for tens of convective times so their mean flows equilibrate. Figure~\ref{fig:w} shows the vertical velocity at the midplane of our highest resolution simulation, R8L8 (see Table~\ref{tab:sims}). The convective up- and down-flows organize themselves into long lanes, which have a horizontal width of about $H$. We define the convective velocity $v_c$ to be the time and volume averaged rms velocity, averaging from an initial time $t_0$ (chosen to remove the influence of initial transients) to the end of the simulation $t_{\rm end}$.
Table~\ref{tab:sims} reports the values of $t_0$ and $t_{\rm end}$, as well as $v_c$, for each simulation. We find that $v_c$ depends on $Ra$, not the domain size, suggesting that increasing the size of the domain from $L_x=Ly=4H$ to $8H$ does not substantially change the (local) convective flow. Furthermore, we expect slightly lower velocities in higher Rayleigh number simulations, because turbulent entrainment slows down the flow.

\begin{figure}
  \centerline{\includegraphics{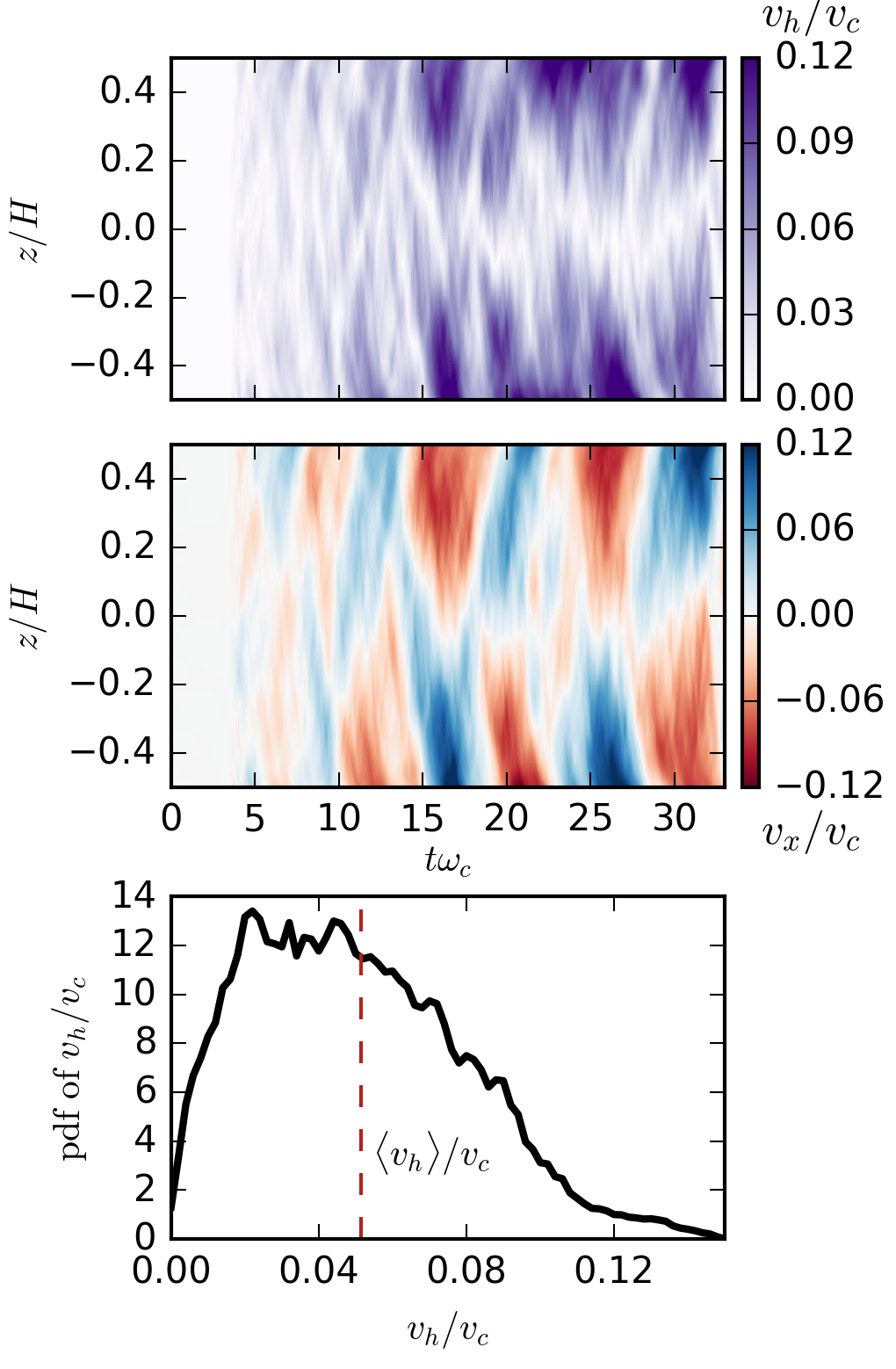}}
  \vspace{-0.3cm}
  \caption{Horizontal velocities in simulation R8L8. The top and middle panel show $v_h/v_c$ (which is positive definite) and $v_x/v_c$ as a function of height and time. The $v_x$ is approximately anti-symmetric across the midplane because the vertical average is zero. The bottom panel shows the pdf of the horizontal velocity, normalized to unity. The mean value (0.053, see Table~\ref{tab:sims}) is denoted with a red dashed line.}
\label{fig:u_mean}
\end{figure}

Finally, we measure the mean horizontal $x$ and $y$ velocities at each time and height $v_x(z,t)$ and $v_y(z,t)$.
We then define $v_h$ to be the magnitude of the velocity in the horizontal plane, i.e., $v_h^2 = v_x^2 + v_y^2$.
In Figure~\ref{fig:u_mean}, we plot $v_h$ and $v_x$ (the latter is a signed quantity, unlike $v_h$, which is positive definite) as a function of $z$ and $t$ in simulation R8L8. The mean flow is highly temporally intermittent. It also is concentrated at the top and bottom of the domain, with only weak mean flows in the center. This is because each convective cell has predominantly vertical flows on the sides, and predominantly horizontal flows on the top and bottom. The plot of $v_x$ is approximately symmetric across the midplane because the vertical average must be zero.   The bottom panel of Figure \ref{fig:u_mean} shows a temporal pdf of the mean horizontal velocity.   The distribution is quite broad, with factor of $\sim 10$ variations in $v_h/v_c$.  

Table~\ref{tab:sims} reports $v_h/v_c$ for each of our simulations. As expected, $v_h/v_c$ decreases by a factor of about two comparing simulations with $L_x/H=L_y/H=8$ to simulations with $L_x/H=L_y/H=4$.
We also find very similar mean horizontal velocities at our two Rayleigh numbers, indicating that the mean horizontal velocities are relatively insensitive to the degree of turbulence within the small range we have probed here. In all four simulations, we find
\begin{align}
\frac{v_h}{v_c} \approx \frac{1}{2} \frac{H}{\sqrt{L_x L_y}}.
\label{eq:vhsim}
\end{align}
This is only a factor of 2 less than the Cartesian analogue of the analytic estimate in equation \ref{eq:vh}.

\begin{figure}
\centering
\includegraphics[width=87mm]{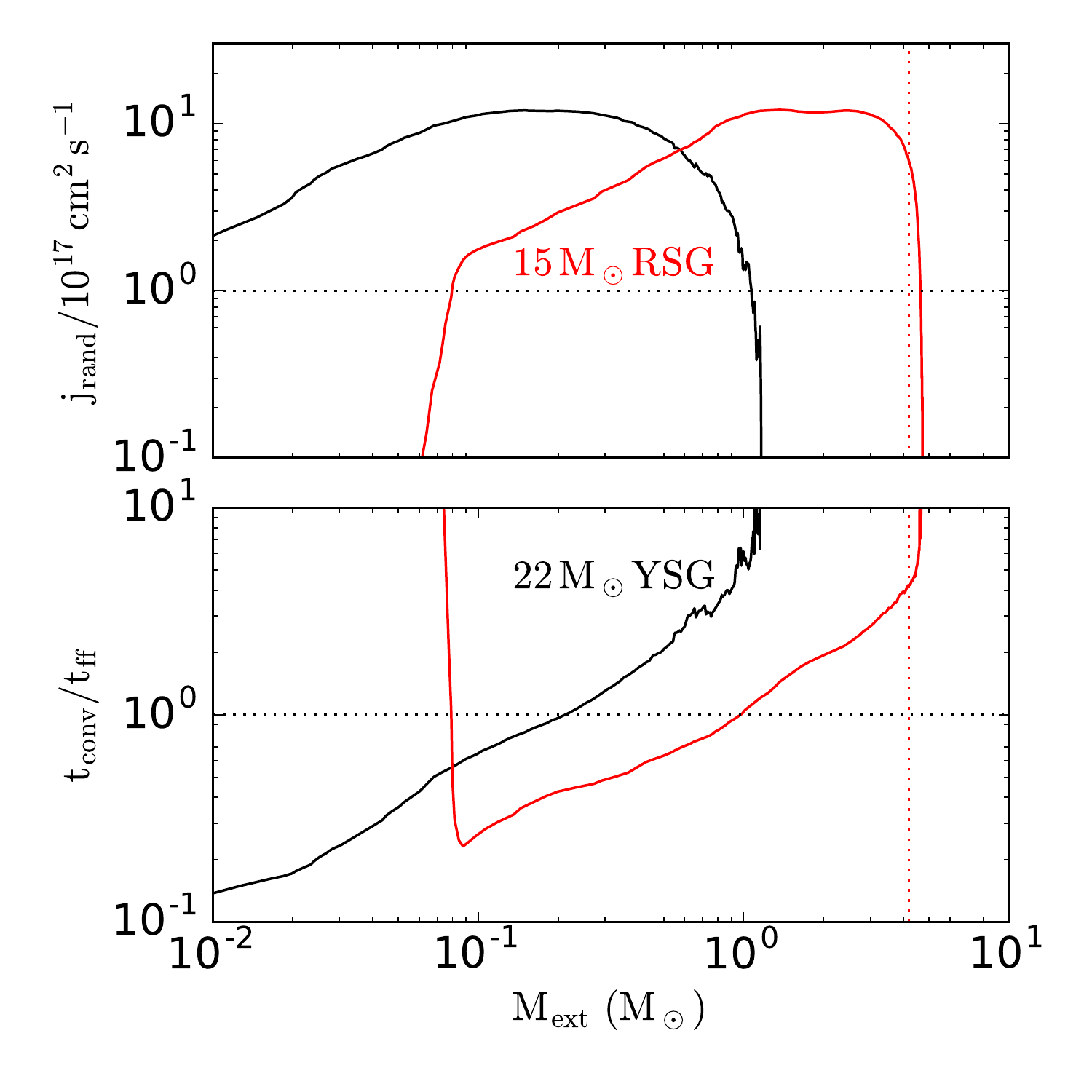} 
\vspace{-0.6cm}
\caption{{\em Top:}. Random angular momentum from mean horizontal convective flows  as a function of exterior mass in solar metallicity pre-SN progenitors (Table \ref{tab}):  a 15 $M_\odot$ red supergiant (red) and a 22 $M_\odot$ yellow supergiant (black).      {\em Bottom:} Ratio of convective turnover time to free-fall time for the same progenitors.   The vertical dotted line indicates the mass coordinate exterior to which the star was unbound by neutrino mass loss in the fiducial simulations of F18.   For the YSG there was very little such unbound mass.
The RSG and YSG have $\sim 0.3$ and $1 \, M_\odot$ of material, respectively, that is not unbound in the failed SN and that has sufficient (random) angular momentum  to form an accretion disc upon collapse.  For much of this material the convection is effectively frozen-in during free-fall because ${\rm t_{conv} \gtrsim t_{ff}}$.}
\label{fig:vhmean}
\end{figure}

\vspace{-0.6cm}
\section{Application to Failed Supernovae }
\label{sec:failedSNe}

Figure \ref{fig:vhmean} (top) shows the random angular momentum $j_{\rm rand}$ produced by mean horizontal convective flows for two pre-SN progenitors:   a 15 $M_\odot$ RSG and a 22 $M_\odot$ YSG.   F18 evolved these progenitors  to core-collapse using {\tt MESA} \citep{Paxton_et_al_2011} (see Table \ref{tab}).     We estimate $j_{\rm rand}$ using equation \ref{eq:jrms}, multiplied by 0.5 per equation \ref{eq:vhsim} and the simulations in \S \ref{sec:sims}; the convective velocities are from mixing length theory.    The random angular momentum decreases at low exterior mass because those regions are close to the stellar surface where the scale-height decreases, suppressing $j_{\rm rand}$ in equation \ref{eq:jrms}; the cutoff in $j_{\rm rand}$ at high exterior mass represents the base of the convection zone near the base of the hydrogen envelope.      Overall, Figure \ref{fig:vhmean} shows that for a significant range of radii/enclosed mass, the random angular momentum from mean flows in convection zones is $\sim 10$ times that associated with the ISCO of a 10 $M_\odot$ black hole, consistent with the analytic expectations from \S \ref{sec:analytics}.   Figure \ref{fig:vhmean} (bottom) shows that for much of this mass the convective turnover time exceeds the free-fall time so that the mean flows present at core-collapse are essentially a random snapshot in time of the convection in the supergiant envelope.   

F18 simulated the collapse of both the RSG and YSG progenitors considered here, along with the weak explosion generated by the change in mass produced by neutrino radiation prior to the proto-NS collapsing to a black hole.   In the RSG progenitor, the outer $\sim 4.2 M_\odot$ of the hydrogen envelope was unbound.   This mass coordinate is indicated by the vertical dotted line in the top panel of Figure \ref{fig:vhmean}.   The mass available to accrete onto the black hole with $j_{\rm rand} \gtrsim j_{ISCO}$ ($M_{\rm acc}$ in Table \ref{tab}) is set by the mass enclosed between the radius $r(M_{\rm ext} = 4.2 M_\odot)$ and the base of the convective envelope.  This is $\sim 0.3 M_\odot$ for the RSG progenitor in Figure \ref{fig:vhmean}.    The core of the YSG progenitor is much more compact than that of the RSG and so the proto-NS quickly collapses to form a black hole.  As a result, the sound pulse and shock generated by neutrino radiation is much weaker than in the RSG and F18 found very little unbound mass.   In this case, nearly all of the convective envelope ($\sim 1 M_\odot$) will accrete onto the newly formed black hole.

A specific angular momentum of $10^{18} j_{18} \, {\rm cm^2 \, s^{-1}}$  (Fig. \ref{fig:vhmean}) will cause the infalling matter to circularize at a radius $r_{\rm circ} \sim 7.5 \, 10^8 \, j_{18}^2 M_{10}^{-1}$ cm $\sim 500  j_{18}^2 M_{10}^{-2} \, GM/c^2$.   This is much smaller than the  radii characteristic of the convective envelope of a supergiant $r_{max} \sim 10^{2-3} R_\odot$ (Table \ref{tab}).    
The ratio of the mass in the disc near the circularization radius  to the mass in the convective envelope is $\sim t_{vis}(r_{\rm circ})/t_{\rm ff}(r_{\rm max})$, the ratio of the viscous time at small radii to the free-fall time at large radii; this is $\ll 1$ given the hierarchy of length-scales involved.   As a result the mass in the disc at any time is a very small fraction of the mass of the convective envelope.   Since the mean horizontal convective flows are coherent over of order a scale-height (Fig. \ref{fig:u_mean}), which is $\sim 0.2-0.4$ of the local radius in the bulk of the convective envelope, this implies that at any time the disc is likely to have a coherent angular momentum set by the particular small region of the convective envelope feeding the disc at that time.    Only on longer timescales that are probably a fraction of $t_{\rm ff}(r_{\rm max})$ (but independent of $r_{\rm circ}$) will the angular momentum of the inflowing material appreciably change magnitude and direction.  

\begin{table*}
\begin{center}
{\bfseries Properties of MESA Pre-Collapse Models}
\end{center}
\begin{center}
\begin{tabular}{lcccccccc}
\hline
 Type & $M_{\rm ZAMS}$  & $M_{\rm cc}$ & $R_{\rm cc}$  &  $\xi_{2.5}$ & $M_{\rm ej}$ & $r_{\rm max}$ & $t_{\rm ff}(r_{\rm max})$ & $M_{\rm acc}$  \\
  \, & ($M_\odot$) & ($M_\odot$) & $(R_{\odot})$ & & $(M_{\odot})$  &  ($R_{\odot}$) & ({\rm days})& ($M_\odot$) \\
\hline
RSG & 15 & 10.8 & 1060 & 0.24 & 4.2 & 200 & 20 & 0.3 \\
YSG & 22 & 11.1 & 690 & 0.54 & $\lesssim 0.1$ & 690 ($\sim R_{cc})$ & 100 & $\sim 1$ \\
\hline
\end{tabular}
\end{center}
\caption{The properties of two solar metallicity {\tt MESA} models collapsed and analyzed in F18:  the type of star, the ZAMS mass $M_{\rm ZAMS}$, the mass of the star at the onset of core collapse $M_{\rm cc}$, the radius of the star at the onset of core collapse $R_{\rm cc}$,  the compactness $\xi_{2.5}$, the mass ejected due to neutrino mass loss alone $M_{\rm ej}$, the radius $r_{\rm max}$ at which the exterior mass in the progenitor is equal to $M_{\rm ej}$ (for the YSG this is the surface of the star), the free-fall time $t_{\rm ff}(r_{\rm max})$ in the progenitor at radius $r_{\rm max}$, and the accreted mass that is rotationally supported due to mean flows in the convection zone $M_{\rm acc}$.} 
\label{tab}
\end{table*}

\vspace{-0.7cm}
\section{Discussion}
\label{sec:discussion}
\vspace{-0.1cm}

We have shown using analytic estimates and 3D  convection simulations that the  finite number of convective eddies in supergiant convection zones leads to appreciable mean horizontal velocities $v_h$ at a given radius in the convection zone, even if the total angular momentum of the star is zero (eqs \ref{eq:jrms} \& \ref{eq:vhsim}; Fig. \ref{fig:u_mean}).   This random angular momentum exceeds that of the inner-most stable circular orbit of a 10 $M_\odot$ black hole by a factor of $\sim 10$ (Fig. \ref{fig:vhmean}).    During the core-collapse of a massive star, the random angular momentum in the convection zone  will lead to the formation of accretion discs in the vicinity of the black hole with ample time for the disc to accrete and generate  outflows that can power an observable transient.    Our analysis neglects the fact that the convection zone will also have radial velocities comparable to the mean horizontal velocities.   This will produce radial mixing during infall, potentially leading to either cancellation or enhancement in the accreted angular momentum.   Collapse simulations beginning from realistic turbulent velocity fields will help better assess this.   It would also be useful to understand the properties of the mean horizontal convective flows in the presence of net rotation since it is likely that the convection will still lead to large variations in the angular momentum of different layers of the stellar envelope.    A final aspect of our model that remains to be explored is how the accretion and outflows at small radii develop given the low binding energy of the infalling material and the fact that at the accretion rates of interest, both photon and neutrino cooling are negligible; we suspect that this makes the disc particularly prone to outflows \citep{Proga2003}.

The convective envelopes of supergiants are weakly bound with binding energies of $\sim 10^{48}$ ergs.   The mechanism that we are proposing for forming black hole accretion discs in core-collapse can thus only work in a failed SN, i.e., when the canonical accretion shock onto the proto-NS fails to promptly revive.    Even in failed SNe,  a weak outgoing shock is produced in many cases, generated by the change in rest mass produced by neutrino radiation from the proto-NS (see \S \ref{sec:intro}). The exact amount of mass ejected in a failed SN depends on the maximum mass of a neutron star and the progenitor structure.    For the specific failed SN models considered here (see Table \ref{tab}; from F18), the YSG  unbinds very little of its convective envelope while the RSG unbinds $\sim 4.2 M_\odot$ (Table \ref{tab}).  However,  an alternative neutrino radiation approximation in F18 led to an ejecta mass for the 15 $M_\odot$ RSG of 4.6 $M_\odot$, which would correspond to essentially all of the convective  envelope being unbound.    It is thus quite possible that failed SNe, if they occur, will have significant diversity, ranging from all to none of the convective hydrogen envelope accreting onto the newly formed black hole (the strongest observational candidate to date - see \citealt{Adams2017} - does not appear to have produced an energetic transient from black hole accretion like that envisioned here).   There will also  be significant diversity in the magnitude of the random angular momentum in the convective envelope.  The reason is that core-collapse happens on of order a free-fall time and thus corresponds to effectively a single snapshot in the history of the turbulent flows in the bulk of the convection zone (Fig. \ref{fig:vhmean} [bottom]).  Figure \ref{fig:u_mean} (bottom panel) shows that there is a factor of $\sim 10$  temporal dispersion in the magnitude of the mean  horizontal convective velocity, which would produce significant differences in the resulting accretion disc at small radii.   

Our  calculations have focused on yellow and red supergiants but the same ideas apply to blue supergiants as well.  We find that the random angular momentum in blue supergiant envelopes is a strong function of the mass of the hydrogen envelope or, equivalently, the stellar radius, with larger hydrogen envelopes more likely to produce rotationally supported material.    However, the large temporal dispersion in the magnitude of $v_h$ in Figure \ref{fig:u_mean} suggests that even some blue supergiants with low mass hydrogen envelopes are likely to form accretion discs via the mechanism proposed here.   Transients associated with blue supergiants will likely be shorter duration and less energetic than those associated with red or yellow supergiants.   Our mechanism will not produce rotationally supported material in compact progenitors like Wolf-Rayet stars.

How would the nominally failed SNe envisioned here manifest observationally?    Previous work has shown that black hole accretion powered outflows and jets can produce a variety of transients, from non-thermal long-duration high energy transients such as ultra-long GRBs and Swift J1644+57 \citep{Quataert_Kasen_2012, Woosley_Heger_2012} to thermal SN light-curves powered by accretion energy \citep{Dexter_Kasen_2013}.    Our primary contribution is to show that for supergiants, significant rotation is not a necessary condition for generating such transients.   This likely expands the range of stellar progenitors prone to producing hydrogen-rich accretion powered transients  (e.g., low metallicity to minimize winds that extract angular momentum is not a prerequisite).    In addition, the fluctuating direction and magnitude of the  angular momentum in the convection zone will produce highly time variable accretion at small radii, with the winds/jet likely covering a much larger solid angle than in a collapse with a well-defined angular momentum direction.   We suspect that this will make it more likely for the accretion energy to unbind the stellar envelope, producing a successful supernova (not unlike the model of \citealt{Papish2011}, although our mechanism for producing the fluctuating jet/wind direction is quite different).   In addition, it is natural to associate the transients predicted here with unusually energetic Type II SNe (e.g., OGLE-2014-SN-073; \citealt{Terreran2017}) and highly time variable sources such as the non-thermal gamma-ray transient Swift J1644+57 (e.g., \citealt{Bloom2011}) or the variable and long-duration Type II supernova iPTF14hls \citep{Arcavi2017}.  Finally, we note that the failed explosion of a blue supergiant has been invoked \citep{Margutti2018} as a possible explanation for the low inferred ejecta mass in SN 2018cow \citep{Prentice2018}.  An accretion disc formed by the mechanism proposed here could also explain the observational evidence for a central engine in this event.

\vspace{-0.7cm}
\section*{Acknowledgements}  We thank Dan Kasen and Rodrigo Fernandez for useful conversations.   This work was supported in part by a Simons Investigator award from the Simons Foundation (EQ) and the Gordon and Betty Moore Foundation through Grant GBMF5076.   ERC was supported by NASA through the Einstein Fellowship Program, Grant PF6-170150.   DL is supported by PCTS and Lyman Spitzer Jr fellowships. Computations were conducted with supported by the NASA High End Computing (HEC) Program through the NASA Advanced Supercomputing (NAS) Division at Ames Research Center on Pleiades with allocation GID s1647.
\vspace{-0.7cm}

\bibliographystyle{mn2e}
\bibliography{ref}

\begin{thebibliography}{}

\bibitem[\protect\citeauthoryear{{Adams}, {Kochanek}, {Gerke}, {Stanek} \&
  {Dai}}{{Adams} et~al.}{2017}]{Adams2017}
{Adams} S.~M.,  {Kochanek} C.~S.,  {Gerke} J.~R.,  {Stanek} K.~Z.,    {Dai} X.,
   2017, \mnras, 468, 4968

\bibitem[\protect\citeauthoryear{{Arcavi}, {Howell}, {Kasen}, {Bildsten} \&
  {Hosseinzadeh}}{{Arcavi} et~al.}{2017}]{Arcavi2017}
{Arcavi} I.,  {Howell} D.~A.,  {Kasen} D.,  {Bildsten} L.,    {Hosseinzadeh} G.
  e.~a.,  2017, \nat, 551, 210

\bibitem[\protect\citeauthoryear{{Bloom}, {Giannios}, {Metzger}, {Cenko} \&
  {Perley}}{{Bloom} et~al.}{2011}]{Bloom2011}
{Bloom} J.~S.,  {Giannios} D.,  {Metzger} B.~D.,  {Cenko} S.~B.,    {Perley}
  D.~A. e.~a.,  2011, Science, 333, 203

\bibitem[\protect\citeauthoryear{{Burns}, {Vasil}, {Oishi}, {Lecoanet} \&
  {Brown}}{{Burns} et~al.}{2016}]{burns16}
{Burns} K.~J.,  {Vasil} G.~M.,  {Oishi} J.~S.,  {Lecoanet} D.,    {Brown} B., ,
  2016, {Dedalus: Flexible framework for spectrally solving differential
  equations}, Astrophysics Source Code Library

\bibitem[\protect\citeauthoryear{{Coughlin}, {Quataert}, {Fern{\'a}ndez} \&
  {Kasen}}{{Coughlin} et~al.}{2018}]{Coughlin18}
{Coughlin} E.~R.,  {Quataert} E.,  {Fern{\'a}ndez} R.,    {Kasen} D.,  2018,
  \mnras, 477, 1225

\bibitem[\protect\citeauthoryear{{Dexter} \& {Kasen}}{{Dexter} \&
  {Kasen}}{2013}]{Dexter_Kasen_2013}
{Dexter} J.,  {Kasen} D.,  2013, ApJ, 772, 30

\bibitem[\protect\citeauthoryear{{Ertl}, {Janka}, {Woosley}, {Sukhbold} \&
  {Ugliano}}{{Ertl} et~al.}{2016}]{Ertl2016}
{Ertl} T.,  {Janka} H.-T.,  {Woosley} S.~E.,  {Sukhbold} T.,    {Ugliano} M.,
  2016, \apj, 818, 124

\bibitem[\protect\citeauthoryear{{Fern{\'a}ndez}, {Quataert}, {Kashiyama} \&
  {Coughlin}}{{Fern{\'a}ndez} et~al.}{2018}]{fernandez18}
{Fern{\'a}ndez} R.,  {Quataert} E.,  {Kashiyama} K.,    {Coughlin} E.~R.,
  2018, \mnras, 476, 2366

\bibitem[\protect\citeauthoryear{{Kochanek}}{{Kochanek}}{2015}]{Kochanek_2015}
{Kochanek} C.~S.,  2015, \mnras, 446, 1213

\bibitem[\protect\citeauthoryear{{Lovegrove} \& {Woosley}}{{Lovegrove} \&
  {Woosley}}{2013}]{Lovegrove_Woosley_2013}
{Lovegrove} E.,  {Woosley} S.~E.,  2013, ApJ, 769, 109

\bibitem[\protect\citeauthoryear{{MacFadyen} \& {Woosley}}{{MacFadyen} \&
  {Woosley}}{1999}]{MacFadyen_Woosley_1999}
{MacFadyen} A.~I.,  {Woosley} S.~E.,  1999, ApJ, 524, 262

\bibitem[\protect\citeauthoryear{{Margutti}, {Metzger}, {Chornock}, {Vurm},
  {Roth}, {Grefenstette} \& {Savchenko}}{{Margutti}
  et~al.}{2018}]{Margutti2018}
{Margutti} R.,  {Metzger} B.~D.,  {Chornock} R.,  {Vurm} I.,  {Roth} N.,
  {Grefenstette} B.~W.,    {Savchenko} V. e.~a.,  2018, ArXiv e-prints

\bibitem[\protect\citeauthoryear{{Moriya}, {Terreran} \& {Blinnikov}}{{Moriya}
  et~al.}{2018}]{Moriya2018}
{Moriya} T.~J.,  {Terreran} G.,    {Blinnikov} S.~I.,  2018, \mnras, 475, L11

\bibitem[\protect\citeauthoryear{{Nadezhin}}{{Nadezhin}}{1980}]{Nadezhin_1980}
{Nadezhin} D.~K.,  1980, Ap\&SS, 69, 115

\bibitem[\protect\citeauthoryear{{O'Connor} \& {Ott}}{{O'Connor} \&
  {Ott}}{2011}]{OConnor_Ott_2011}
{O'Connor} E.,  {Ott} C.~D.,  2011, \apj, 730, 70

\bibitem[\protect\citeauthoryear{{Papish} \& {Soker}}{{Papish} \&
  {Soker}}{2011}]{Papish2011}
{Papish} O.,  {Soker} N.,  2011, \mnras, 416, 1697

\bibitem[\protect\citeauthoryear{{Paxton}, {Bildsten}, {Dotter}, {Herwig},
  {Lesaffre} \& {Timmes}}{{Paxton} et~al.}{2011}]{Paxton_et_al_2011}
{Paxton} B.,  {Bildsten} L.,  {Dotter} A.,  {Herwig} F.,  {Lesaffre} P.,
  {Timmes} F.,  2011, \apjs, 192, 3

\bibitem[\protect\citeauthoryear{{Prentice}, {Maguire}, {Smartt}, {Magee},
  {Schady} \& {Sim}}{{Prentice} et~al.}{2018}]{Prentice2018}
{Prentice} S.~J.,  {Maguire} K.,  {Smartt} S.~J.,  {Magee} M.~R.,  {Schady} P.,
     {Sim} S. e.~a.,  2018, \apjl, 865, L3

\bibitem[\protect\citeauthoryear{{Proga} \& {Begelman}}{{Proga} \&
  {Begelman}}{2003}]{Proga2003}
{Proga} D.,  {Begelman} M.~C.,  2003, \apj, 592, 767

\bibitem[\protect\citeauthoryear{{Quataert} \& {Kasen}}{{Quataert} \&
  {Kasen}}{2012}]{Quataert_Kasen_2012}
{Quataert} E.,  {Kasen} D.,  2012, MNRAS, 419, L1

\bibitem[\protect\citeauthoryear{{Spiegel} \& {Veronis}}{{Spiegel} \&
  {Veronis}}{1960}]{spiegel60}
{Spiegel} E.~A.,  {Veronis} G.,  1960, \apj, 131, 442

\bibitem[\protect\citeauthoryear{{Sukhbold}, {Woosley} \& {Heger}}{{Sukhbold}
  et~al.}{2018}]{Sukhbold2018}
{Sukhbold} T.,  {Woosley} S.~E.,    {Heger} A.,  2018, \apj, 860, 93

\bibitem[\protect\citeauthoryear{{Terreran}, {Pumo}, {Chen}, {Moriya} \&
  {Taddia}}{{Terreran} et~al.}{2017}]{Terreran2017}
{Terreran} G.,  {Pumo} M.~L.,  {Chen} T.-W.,  {Moriya} T.~J.,    {Taddia} F.
  e.~a.,  2017, Nature Astronomy, 1, 713

\bibitem[\protect\citeauthoryear{{Woosley} \& {Heger}}{{Woosley} \&
  {Heger}}{2012}]{Woosley_Heger_2012}
{Woosley} S.~E.,  {Heger} A.,  2012, ApJ, 752, 32

\end{thebibliography}


\end{document}